\documentclass[11pt,bezier]{article}
\usepackage{amsmath,amssymb,amsfonts,euscript}
\usepackage{algorithmic, algorithm}

\usepackage{graphicx}

\usepackage[ansinew]{inputenc}
\usepackage{graphicx}
\usepackage{color}
\usepackage{mathrsfs}

\usepackage[colorlinks]{hyperref}
\usepackage[active,new,noold,marker]{xrcs}
\catcode`\=13
\def{$\bowtie$}

\usepackage{eurosym}
\usepackage{lscape} 

\newtheorem{prethm}{{\bf Theorem}}

\newenvironment{thm}{\begin{prethm}{\hspace{-0.5
               em}{\bf}}}{\end{prethm}}

\newtheorem{prepro}{{\bf Theorem}}

\newenvironment{pro}{\begin{prepro}{\hspace{-0.5
               em}{\bf}}}{\end{prepro}}

\newtheorem{preprop}{{\bf Proposition}}

\newtheorem{precor}{{\bf Corollary}}

\newtheorem{preconj}{{\bf Conjecture}}

\newtheorem{predefi}{{\bf Definition}}

\newenvironment{defi}{\begin{predefi}{\hspace{-0.5
               em}{\bf}}}{\end{predefi}}

\newtheorem{preremark}{{\bf Remark}}

\newenvironment{remark}{\begin{preremark}\rm{\hspace{-0.5
               em}{\bf}}}{\end{preremark}}

\newtheorem{preexample}{{\bf Fact}}

\newtheorem{prelem}{{\bf Lemma}}

\newtheorem{prelam}{{\bf Lemma}}

\newtheorem{preprob}{{\bf Problem}}

\newenvironment{prob}{\begin{preprob}{\hspace{-0.5
               em}{\bf.}}}{\end{preprob}}

\newtheorem{preproof}{{\bf Proof}}

\newtheorem{preali}{{\bf Proof of Theorem 1.}}

\newtheorem{prealii}{{\bf Proof of Theorem 2.}}

\newtheorem{prealiii}{{\bf Proof of Theorem 3.}}

\title{  Algorithmic Complexity of Proper Labeling Problems}

\author{{\normalsize
{\sc Ali Dehghan${}^{\mathsf{a}}$},\,
{\sc Mohammad-Reza Sadeghi${}^{\mathsf{a}}$},\,
 {\sc Arash Ahadi${}^{\mathsf{b}}$}\,
}\vspace{3mm}
\\{\footnotesize{${}^{\mathsf{a}}$\it Faculty of Mathematics and Computer Science,
Amirkabir University of Technology, Tehran,
Iran}}  {\footnotesize{}}\\{\footnotesize{${}^{\mathsf{b}}$\it
Department of
Mathematical Sciences, Sharif University of Technology, Tehran, Iran}}
\thanks{{\it E-mail addresses}:  $\mathsf{ali\_dehghan16@aut.ac.ir}$, $\mathsf{msadeghi@aut.ac.ir}$, $\mathsf{arash\_ahadi@mehr.sharif.edu}$. } }

\date{}
\begin{document}
\maketitle

\begin{abstract}
{\small \noindent
A proper labeling of a graph is an assignment of integers to some elements of a graph, which
may be the vertices, the edges, or both of them, such that  we obtain a proper vertex coloring via the labeling subject to
some conditions. The problem of proper labeling offers many variants and received a great interest during recent
years. We  consider the algorithmic complexity of some variants of the  proper labeling problems, we  present some polynomial time algorithms and $ \mathbf{NP} $-completeness results for them.

}

\begin{flushleft}
\noindent {\bf Key words:}  Proper labeling; Computational complexity; Multiplicative vertex-coloring weightings; Gap vertex-distinguishing edge colorings ; Fictional coloring; Vertex-labeling by maximum;
1, 2, 3-Conjecture; Multiplicative
1, 2, 3-Conjecture.

\noindent {\bf Subject classification: 05C15, 05C20, 68Q25}

\end{flushleft}

\end{abstract}

\section{Introduction}
\label{}

A proper labeling of a graph is an assignment of integers to some elements of a graph, which
may be the vertices, the edges, or both of them, such that  we obtain a proper vertex coloring via the labeling subject to
some conditions.
Karo\'nski, \L{}uczak and Thomason
initiated the study of proper labeling \cite{MR2047539}. They introduced an edge-labeling which
is additive vertex-coloring that means for every edge
$uv$, the sum of labels of the edges incident to $u$ is different
from the sum of labels of the edges incident to $v$ \cite{MR2047539}.
The problem of proper labeling offers many variants and received a great interest during recent
years, for instance see \cite{MR2145514, MR2729020,  MR2552893,  MR2595676, MR2047539, MR2428690}.

There is a concept which has a close relationship with proper labeling problem.
An {\it adjacent vertex-distinguishing edge coloring}, or {\it avd-coloring}, of a graph $G$ is a proper edge coloring of $G$ such that no pair of adjacent vertices meets the same set of colors. Avd-coloring has been widely studied, for example see \cite{MR2299707, MR1469354, MR2171364}.

In this paper, we consider the algorithmic complexity of  the following  proper labeling problems.
We will present some polynomial time algorithms and $ \mathbf{NP} $-completeness results for them.
Throughout the paper we denote $\{1,2,\cdots, k\}$ by $\mathbb{N}_k$.

{{\bf    Edge-labeling by sum}}.\\
An edge-labeling $f$ is an {\em  edge-labeling by sum} if $c(v)=\sum_{e \ni v} f(e)$  is a proper vertex coloring. This parameter was introduced by Karo\'nski et al. and it is conjectured that three labels
$\{1,2,3\}$ are sufficient for every connected graph, except
$K_2$ (1, 2, 3-Conjecture, see \cite{MR2047539}).
 This labeling has been studied extensively by several
authors, for instance see \cite{MR2145514,   MR2404230, z2, z3, D2, MR2428690}.
 Currently, we know that  every connected graph has an {\em  edge-labeling by sum}, using the labels from $\mathbb{N}_{5}$ \cite{MR2595676}. Furthermore, Addario-Berry, Dalal, and Reed  showed that almost all graphs have an {\em  edge-labeling by sum} from $\mathbb{N}_{2}$ \cite{MR2404230}.

Dudek and Wajc
 showed that determining whether a given graph has an
{\em  edge-labeling by sum} from $ \mathbb{N}_{2}$  is
$ \mathbf{NP} $-complete \cite{David}.
They showed that determining whether a graph has an {\em  edge-labeling by sum} from either $\{0,1\}$ or
$\{1,2\}$ is $ \mathbf{NP} $-complete.
They conjectured that deciding whether a given graph has an {\em edge-labeling by sum} from $\{a,b\}$ for any rational $a$ and $b$ is $ \mathbf{NP} $-complete.
By a short proof we improve the current results and show that for a given 3-regular graph $G$, determining whether $G$ has an
{\em  edge-labeling by sum} from $\{a,b\}$ is $ \mathbf{NP} $-complete.

\begin{thm}\label{pp0}
For a given 3-regular graph $G$, determining whether $G$ has an
{\em  edge-labeling by sum} from $ \mathbb{N}_{2}$ is $ \mathbf{NP} $-complete.
\end{thm}

{{\bf    Vertex-labeling by sum}} (Lucky labeling and sigma coloring).\\
A vertex-labeling $f$ is a {\em  vertex-labeling by sum} if  $c(v)=\sum_{u \sim v} f(u)$ is a proper vertex coloring. {\em  Vertex-labeling by sum} is a vertex
version of the above problem, which was introduced recently by Czerwi\'nski et al. \cite{MR2552893}. It was conjectured that every graph $G$ has a {\em  vertex-labeling by sum}, using the labels  $\{1, 2, \ldots,
 \chi(G) \}$ \cite{MR2552893} and it was shown that every graph $G$ with $\Delta(G)\geq 2$, has a {\em  vertex-labeling by sum}, using the labels $\{1,2, \ldots, \Delta ^2-\Delta +1\}$ \cite{akbari2}.
A similar version of this labeling was introduced by Chartrand et al.  \cite{MR2729020}.

It was shown that, it is $ \mathbf{NP} $-complete to decide for a given planar 3-colorable graph $G$, whether $G$ has a {\em  vertex-labeling by sum} from $ \mathbb{N}_{2}$ \cite{ahadi}. Furthermore, it is $ \mathbf{NP} $-complete to determine  for a 3-regular graph $G$, whether $G$ has a {\em  vertex-labeling by sum} from  $ \mathbb{N}_{2}$ \cite{ali}.

 {{\bf    Edge-labeling by product.}} (Multiplicative vertex-coloring) \\
An edge-labeling $f$ is an {\em  edge-labeling by product} if $c(v)=\prod_{e \ni v} f(e) $  is a proper vertex coloring.
This variant was introduced by Skowronek-Kazi{\'o}w and it
is conjectured that every non-trivial graph $G$ has an
{\em  edge-labeling by product}, using the labels from $\mathbb{N}_3$ (Multiplicative
1, 2, 3-Conjecture, see \cite{product}).
Currently, we know that  every non-trivial  graph has an {\em  edge-labeling by product}, using the labels from $\mathbb{N}_4$ \cite{product}. Also, every non-trivial, 3-colorable graph $G$ permits
an {\em  edge-labeling by product} from $ \mathbb{N}_{3}$ \cite{product}.
We will prove the following for planar 3-colorable graphs and 3-regular graphs.

\begin{thm}\label{pp1} We have the following:\\
{\em (i)} For a given planar 3-colorable graph $G$, determining whether $G$ has an
{\em  edge-labeling by product} from $ \mathbb{N}_{2}$ is $ \mathbf{NP} $-complete.\\
{\em (ii)} For a given 3-regular graph $G$, determining whether $G$ has an
{\em  edge-labeling by product} from $ \mathbb{N}_{2}$ is $ \mathbf{NP} $-complete.
\end{thm}

{{\bf   Vertex-labeling by product.}}\\
A vertex-labeling $f$ is a {\em  vertex-labeling by product} if  $c(v)=\prod_{u \sim v} f(u) $  is a proper vertex coloring.
{\em  Vertex-labeling by product} is a vertex
version of the above problem.

For a given graph $G$, let $g:V(G)\rightarrow \{1,\ldots, \chi(G)\}$ be a proper vertex coloring  of $G$.
Label the set of vertices  $g^{-1}(1)$ by 1; also, for each $i$, $1 <i \leq \chi(G)$ label the set of vertices  $g^{-1}(i)$ by the $(i-1)$-th prime number;
this labeling is a {\em  vertex-labeling by product}.
In number theory,  Prime Number Theorem  describes the asymptotic distribution of the prime numbers.  As a consequence of  Prime Number Theorem we have the following  bound for the size of the $n$-th prime number $p_n$ (i.e., $p_1=2$, $p_2=3$, etc. ):
$p_n < n\ln n +n\ln \ln n$, for $n\geq 6$ (see \cite{PNT} p. 233).
So,  every graph $G$ has a {\em  vertex-labeling by product}, from  $\{1, 2, \ldots,
 \chi \ln \chi  +\chi \ln \ln \chi + 2  \}$.
Here, we ask the following question.

\begin{prob}{
Does every graph $G$ have a {\em  vertex-labeling by product}, using the labels  $\{1, 2, \ldots,
 \chi(G) \}$?
}\end{prob}

Every bipartite graph and  3-colorable graph has a
{\em  vertex-labeling by product} from $ \mathbb{N}_{2}$ and  $ \mathbb{N}_{3}$, respectively.
We will prove the following about the planar 3-colorable graphs.

\begin{thm}\label{pp2}
We have the following:\\
{\em (i)} For a given planar 3-colorable graph $G$, determining whether $G$ has a
{\em  vertex-labeling by product} from $ \mathbb{N}_{2}$ is $ \mathbf{NP} $-complete.\\
{\em (ii)} For every $k$, $k\geq 3$, it is $ \mathbf{NP} $-complete to determine whether a given graph has a
{\em  vertex-labeling by product} from $ \mathbb{N}_k$.
\end{thm}

Also, we prove the following about random graphs.

\begin{thm}\label{random}
For every constant $p$, $p\in (0,1)$, almost all graphs in $G(n, p)$ have a
{\em  vertex-labeling by product} from $ \mathbb{N}_{11}$.
\end{thm}

Let $G$ be a 3-regular graph. $G$ has a  {\em  vertex-labeling by product} from $ \mathbb{N}_{2}$ if and only if
$G$ has a {\em  vertex-labeling by sum} from $ \mathbb{N}_{2}$.
It was proved that it is $ \mathbf{NP} $-complete to determine  for a given 3-regular graph $G$, whether $G$ has a {\em  vertex-labeling by sum} from  $ \mathbb{N}_{2}$ \cite{ali}. So we have the following:

\begin{pro}
It is $ \mathbf{NP} $-complete to determine  for a given 3-regular graph $G$, whether $G$ has a {\em  vertex-labeling by product} from  $ \mathbb{N}_{2}$.
\end{pro}


 {{\bf    Edge-labeling by gap.}}\\
An edge-labeling $f$ is an {\em  edge-labeling by gap} if

$c(v)=\begin{cases}
    1                    & $if$ \,\,d(v)=0,\\
    f(e)_{e \ni v}       & $if$ \,\,d(v)=1,\\
    \max_{e \ni v} f(e) - \min_{e \ni v} f(e)       & $otherwise$,\
\end{cases}$

is a proper vertex coloring.
Every graph $G$ has an {\em edge-labeling by gap} if and only if it has no connected component isomorphic to  $K_2$; for instance, put the different powers of
two on the edges of $G$.

A similar definition was introduced by Tahraoui et al. \cite{gap}.
They introduced the following variant: Let $G$ be a graph, $k$ be a positive integer and $f$ be a mapping from $E(G)$ to the set $\mathbb{N}_{k}$. For each vertex
$v$ of $G$, the label of $v$ is defined as

$c(v)=\begin{cases}
    f(e)_{e \ni v}       & $if$ \,\,d(v)=1,\\
    \max_{e \ni v} f(e) - \min_{e \ni v} f(e)       & $otherwise$,\
\end{cases}$

The mapping $f$ is called {\it gap vertex-distinguishing labeling} if distinct vertices have distinct labels. Such a coloring is called a {\it gap-k-coloring} and is denoted by $gap(G)$ \cite{gap}. It was  conjectured that for a connected graph $G$ of order $n$ with $n>2$, $gap(G)\in \{n-1,n,n+1\}$ \cite{gap}. They propose
 study of the variant of the gap coloring problem that distinguishes the adjacent vertices only.

Now, consider the following example.
\begin{remark}{
Every  complete graph $K_n$ of order $n$ with $n>2$, has an {\em  edge-labeling $f_n$ by gap} from $\{1,2, \cdots, \chi(K_n)+1\}$.
Suppose that $K_3 = v_1v_2v_3$ and let $f_3$ be the following
function: $f_3(v_1v_2) = 4$, $f_3(v_1v_3) = 1$ and $f_3(v_2v_3) = 2$. Define $f_n$ recursively:

$f_n(v_i v_j)=\begin{cases}
    f_{n-1}(v_i v_j)+1       & $if $ i,j<n,\\
    1                       & $if $( i=n $ and $ j\neq 2 )$ or $ ( j=n $ and $ i\neq 2 ),\\
    2       & $otherwise$.\
\end{cases}$
}\end{remark}

Let $f$ be an {\em  edge-labeling by gap} from $\mathbb{N}_{k}$, for a graph $G$ we have $k\geq \chi(G)-1$.
Now, we state the following problem:

\begin{prob}{
Does every connected graph $G$ of order $n$ with $n>2$, have an {\em  edge-labeling by gap} from $\{1,2, \ldots, \chi(G)+1 \}$?
}\end{prob}

We will prove the following:

\begin{thm}\label{pp3}
We have the following:\\
{\em (i)} For a given planar bipartite graph $G$ with minimum degree two, determining whether $G$ has an
edge-labeling by gap from $ \mathbb{N}_{2}$ is in $ \mathbf{P} $.\\
{\em (ii)} For a given planar bipartite graph $G$, determining whether $G$ has an
edge-labeling by gap from $ \mathbb{N}_{2}$ is $ \mathbf{NP} $-complete.\\
{\em (iii)} For every $k$, $k\geq 3$, it is $ \mathbf{NP} $-complete to determine whether a given graph has an
{\em  edge-labeling by gap} from $ \mathbb{N}_k$.
\end{thm}


 {{\bf    Vertex-labeling by gap.}}\\
A vertex-labeling $f$ is a {\em  vertex-labeling by gap} if

$c(v)=\begin{cases}
                   1       & $if$ \,\,d(v)=0,\\
    f(u)_{u \sim v}       & $if$ \,\,d(v)=1,\\
    \max_{u \sim v} f(u) - \min_{u \sim v} f(u)       & $otherwise$,\
\end{cases}$

is a proper vertex coloring. A graph may lack any {\em  vertex-labeling by gap}.
Here we ask the following:

 \begin{prob}{
Does there exist a polynomial time algorithm to determine whether a given graph  has a {\em  vertex-labeling by gap}?
}\end{prob}

Every  tree $T$  has a {\em  vertex-labeling by gap} from $ \mathbb{N}_{2}$.
Let $T$ be a tree and $v\in V(T)$ be an arbitrary vertex, define:

$f(u)=\begin{cases}
    1       & $if$ \,\,d(u,v)\equiv 0 \ (\mod 4),\\
    2       & $otherwise$.\
\end{cases}$

It is easy to see that this labeling is a
{\em  vertex-labeling by gap} from $ \mathbb{N}_{2}$.

We will show that there is dichotomy for the problem determining whether a  given graph has a
{\em vertex-labeling by gap} from $ \mathbb{N}_{2}$: it is
polynomially solvable for planar bipartite graphs but $ \mathbf{NP} $-complete for  bipartite graphs.

\begin{thm}\label{pp4}
We have the following:\\
{\em (i)} For a given planar bipartite graph $G$, determining whether $G$ has a
{\em vertex-labeling by gap} from $ \mathbb{N}_{2}$ is in $ \mathbf{P} $.\\
{\em (ii)} For a given  bipartite graph $G$, determining whether $G$ has a
vertex-labeling by gap from $ \mathbb{N}_{2}$ is $ \mathbf{NP} $-complete.\\
{\em (iii)} For every $k$, $k\geq 3$, it is $ \mathbf{NP} $-complete to determine whether a given graph has a
{\em   vertex-labeling by gap} from $ \mathbb{N}_k$.\\
{\em (iv)} It is $ \mathbf{NP} $-complete to
decide whether a given  planar 3-colorable graph $G$ has a
{\em vertex-labeling by gap} from $ \mathbb{N}_{2}$.
\end{thm}


Note that, every bipartite graph $G=[X,Y]$ has a {\em  vertex-labeling by gap}, label the set of vertices $X$ by 1
 and label the set of vertices   $Y$ by different even numbers.

\begin{remark}{
A hypergraph $H$ is a pair $(X, Y)$,
where $X$ is the set of vertices and $Y$ is a set of non-empty subsets of $X$, called edges. The $k$-coloring of $H$ is a coloring $\ell:X \rightarrow \mathbb{N}_{k}$ such
that, for every edge $e$ with $\mid e\mid > 1$, there exist $v,u\in X$ such that $\ell(u) \neq \ell(v)$.
It was shown by Thomassen \cite{MR1135027}
that, for any $k$-uniform and $k$-regular hypergraph $H$, if $k \geq 4$, then $H$ is 2-colorable.
For a given $r$-regular bipartite graph $G=[X,Y]$ with $r>3$, consider the hypergraph $H$ with the vertex set $X$ and edge set $Y$ such that such that  $v \in e$ in $H$  if and only if $v\in  X $ is adjacent to $e\in  Y $ in $G$. Let $\ell$ be a
$2$-coloring for $H$. Define

$c(v)=\begin{cases}
    \ell(v)       & $if$ \,\,v\in X,\\
    2       & $otherwise$,\
\end{cases}$

this labeling is {\em  vertex-labeling by gap} from $ \mathbb{N}_{2}$.
So every $r$-regular bipartite graph $G=[X,Y]$ with $r\geq 4$, has a
{\em  vertex-labeling by gap} from $ \mathbb{N}_{2}$.
But, there are $3$-uniform, $3$-regular hypergraphs that are not $2$-colorable. For instance, consider the Fano Plane. The Fano Plane is a hypergraph  with seven vertices $ \mathbb{Z}_{7} $ and seven edges $ \{ \{ i,i+1,i+3 \}: 1 \leq i \leq 7\}$.
}\end{remark}

 \begin{prob}{
Determine the computational complexity of deciding whether a given 3-regular bipartite graph $G$ have a
{\em  vertex-labeling by gap} from $ \mathbb{N}_{2}$.
}\end{prob}

{{\bf    Vertex-labeling by degree.}} (Fictional coloring)\\
A vertex-labeling $f$ is a {\em  vertex-labeling by degree} if  $c(v)= f(v)d(v)$, where $d(v)$ is the degree of vertex $v$, is a proper vertex coloring. This parameter was introduced by Bosek, Grytczuk, Matecki and \.{Z}elazny \cite{personal}. They
conjecture that every graph $G$ has a {\em  vertex-labeling by degree} from $\{1, 2, \cdots, \chi(G)\}$. Let $p$ be a prime number and let $G$ be a graph such that  $\chi(G)\leq p -1$, they proved that
$G$ has a {\em  vertex-labeling by degree} from $ \mathbb{N}_{p-1}$, so
every graph $G$ has a {\em  vertex-labeling by degree} from $\{1, 2, \cdots, 2\chi(G)\}$ \cite{personal}.

Since determining the chromatic number of regular graph is  $ \mathbf{NP}$-hard, hence for
a given regular graph $G$,
determining the minimum
$k$ such that $G$ has a
{\em  vertex-labeling by degree} from $  \mathbb{N}_k $ is $ \mathbf{NP} $-hard.
We will prove the following:

\begin{thm}\label{pp5}\\
{\em (i)} Determining whether a given  graph has a
{\em  vertex-labeling by degree} from $\mathbb{N}_{2}$ is in  $ \mathbf{P}$.\\
{\em (ii)} For every $k$, $k\geq 3$, for a given  graph $G$ with $\chi(G)=k+1$ determining wether $G$ has
{\em  vertex-labeling by degree} from $\mathbb{N}_k$ is $ \mathbf{NP} $-complete.
\end{thm}

{{\bf    Vertex-labeling by maximum.}}\\
A vertex-labeling $f$ is a {\em  vertex-labeling by maximum} if  $c(v)=\max_{u \sim v} f(u) $ is a proper vertex coloring.
A graph $G$ may lack any {\em  vertex-labeling by maximum} and
 it has a
{\em  vertex-labeling by maximum} from $ \mathbb{N}_{2}$ if and only if $G$ is bipartite.

\begin{remark}{
Let $k$ be the minimum number such that $G$ has a {\em  vertex-labeling by maximum} from the set $\mathbb{N}_k$, then $k-\chi(G)$ can be arbitrarily large.
For instance, for a given $t$, $t> 3$ consider the graph $G$ with vertex set $V(G)=\{ a_i : 1\leq i \leq t\}\cup \{b_j: 1 \leq j \leq t-2\}$ and edge set $E(G)=\{   a_i a_{i+1} : 1\leq i \leq t-1\}\cup \{   a_j b_{j},b_{j}a_{j+1} : 1\leq j \leq t-2\}$. $G$ is 3-colorable but $k=t$.
}\end{remark}

Every triangle-free graph has a {\em  vertex-labeling by maximum} (put different numbers on vertices) and if $G$ is a graph such that every vertex appears in some triangles, then $G$ does not have any {\em  vertex-labeling by maximum}.
Here, we present a nontrivial necessary condition that can be checked in polynomial time for a graph to have a {\em  vertex-labeling by maximum}. First consider the following definition.

\begin{defi}{ For a given graph $G$ the subset $S$ of vertices is called {\it triangular-structured-vertices (TSV)} if every $v\in S$ appears in a triangle in $G[S]$ and for every two adjacent vertices $v$ and $u$, where $v\in S$ and $u \notin S$, there exists a vertex $z\in S$ such that $z$ is adjacent to $v$ and $u$.
}\end{defi}

Let $S $ be a TSV for $G$. By way of contradiction, assume  that $f$ is a {\em  vertex-labeling by maximum} for $G$ and $v\in S \cup N(S)$ is a vertex such that $f(v)=\max_{u \in S \cup N(S)} f(u)$. Then  $v$ has two neighbors $x$ and $y$ in $S$ with $\max_{u \sim x } f(u)=\max_{u \sim y } f(u)=f(v)$. This is a contradiction. Therefore, if $G$ has a TSV, then $G$ does not have a {\em  vertex-labeling by maximum}.

\begin{thm}\label{pp6}
We have the following:\\
{\em (i)} For a given  3-regular graph $G$, determining whether $G$ has a
{\em  vertex-labeling by maximum} from $ \mathbb{N}_{3}$ is $ \mathbf{NP} $-complete.\\
{\em (ii)} For every $k$, $k\geq 3$, it is
{\bf NP}-complete to decide whether $G$ has a
{\em  vertex-labeling by maximum} from $ \mathbb{N}_k$ for a given $k$-colorable graph
$G$.\\
{\em (iii)} There is a polynomial time algorithm to determine whether $G$ has a TSV.
\end{thm}

 Here, we ask the following questions:

 \begin{prob}{
Is the necessary condition, sufficient for a
given graph to have a {\em  vertex-labeling by maximum}?
}\end{prob}

 \begin{prob}{
Does there exist a polynomial time algorithm to determine whether a given graph  has a {\em  vertex-labeling by maximum}?
}\end{prob}

{{\bf Notation}}

Throughout this paper all graphs are finite and simple.
We follow \cite{MR1567289, MR1367739} for terminology and
notations are not defined here, and we consider finite undirected
simple graphs $G=(V(G),E(G))$.
We denote a 3-Sat formula by $\Phi=(X,C) $, where $X$ is the set  of variables and $C$ is a set  of clauses over $X$
 such that each   clause $c \in C$ has $\mid c  \mid = 3$.

A {\it proper vertex coloring} of $G$ is a function $c:
V(G)\longrightarrow L$, such that if $u,v\in V(G)$ are adjacent,
then $c(u)$ and $c(v)$ are different.
A {\it proper vertex $k$-coloring}
is a proper vertex coloring with $|L|=k$.
The smallest integer $k$ such that
$G$ has a proper vertex $k$-coloring is called the {\it chromatic number} of $G$ and denoted by $\chi(G)$.
Similarly, for $ k\in \mathbb{N} $, a {\it proper edge $k$-coloring} of $G$ is a function $c:
E(G)\longrightarrow \mathbb{N}_k$, such that if $e,e'\in E(G)$ share a common endpoint,
then $c(e)$ and $c(e')$ are different.
The smallest integer $k$ such that
$G$ has a proper edge $k$-coloring is called the {\it edge chromatic number} of $G$ and is denoted by $\chi '(G)$. By Vizing's theorem \cite{MR0180505}, the edge chromatic number of a graph $G$ is equal to either $ \Delta(G) $ or $ \Delta(G) +1 $. Those graphs
$G$ for which $\chi '(G)=\Delta(G)  $ are said to belong to $Class$ $1$, and the others to $Class$ $2$.

We denote the induced subgraph $G$ on $S$ by $G[S]$.
Also, for every $v\in V (G)$ and $S \subseteq V(G)$, $N(v)$ and $N(S)$  denote  the neighbor set of $v$ and the set of vertices of $G$ which has a neighbor in $S$, respectively.
Furthermore, let $H$ be a subgraph of $G$, for every vertex $v$, $v \in V(H)$, we denote $\mid \{ u : vu \in E(H), u \in V(H)\} \mid$ by $d_{H}(v)$.

{{\bf Summery of results}}

\begin{table}[ht]
\caption{Recent results on proper labeling of graphs } 
\centering 
\begin{tabular}{l c  c c c} 
\hline\hline 
Edge  &  $\mathbf{NP} $-h  & $\mathbf{P} $   &   Upper Bound &  Conjecture \\ [0.5ex] 
\hline 
Sum &     3-regular    & - & $\mathbb{N}_{5}$ & $\mathbb{N}_{3}$  \\ \\
Product & 3-regular     & - &$\mathbb{N}_{4}$ & $\mathbb{N}_{3}$\\
         &       Planar 3-col     &                  &              \\ \\
Gap &   Planar bipartite  & Planar bipartite with $\delta>1$ & - &  $\mathbb{N}_{ \chi +1}  $\\
\hline
Vertex  &   &       &  \\ [0.5ex] 
\hline
Sum &  3-regular   & - & $\mathbb{N}_{\Delta ^2-\Delta +1}$ & $\mathbb{N}_{\chi  }$\\
     &      Planar 3-col    &                  &              \\ \\
Product & 3-regular  & -  & $\mathbb{N}_{\chi \ln \chi  +\chi \ln \ln \chi + 2 }   $ &  $\mathbb{N}_{\chi  }$\\
         &       Planar 3-col     &                  &              \\ \\
Degree &   4-col  & From $\mathbb{N}_{2}$ & $\mathbb{N}_{2\chi }$ & $\mathbb{N}_{\chi  }$\\ \\
Maximum &  3-regular   & Bipartite & - & - \\ \\
Gap & Bipartite    &Planar Bipartite & - & - \\
         &       Planar 3-col     &                  &              \\
 [1ex] 
\hline 
\end{tabular}
\label{table:nonlin} 
\end{table}

\section{Proof of Theorem \ref{pp0}}
We reduce {\em   Monotone Not-All-Equal (NAE) 3-Sat} to our problem in polynomial time. It is shown that the following problem is $ \mathbf{NP}$-complete \cite{MR1567289}.

 {\em Monotone Not-All-Equal 3-Sat .}\\
\textsc{Instance}: A 3-Sat formula $(X,C)$ such that there is no negation in the formula.\\
\textsc{Question}: Is there a truth assignment for $X$ such that each clause in $C$ has at
least one  true literal and at least one false literal?

\begin{figure}[ht]
\begin{center}
\includegraphics[scale=.7]{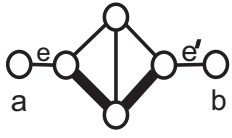}
\caption{The  gadgets $B_x$. In every {\em  edge-labeling $f$ by sum} from $ \mathbb{N}_{2}$, $f(e)=f(e')$.} \label{TBT}
\end{center}
\end{figure}

\begin{figure}[ht]
\begin{center}
\includegraphics[scale=.4]{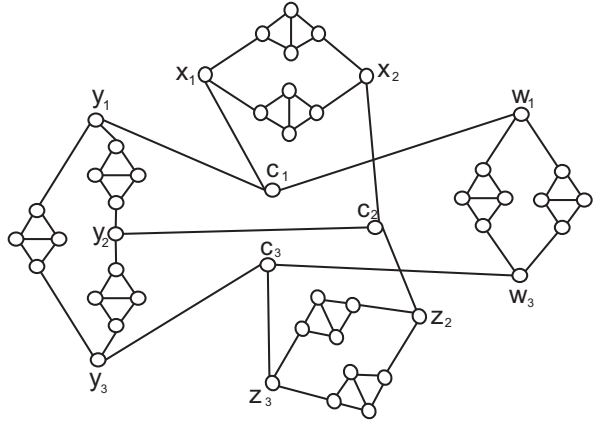}
\caption{The graph $G$ derived from the   formula $ \Phi= c_1 \wedge c_2 \wedge c_3$, where
$ c_1 =  x \vee y \vee w$, $ c_2 =x \vee y \vee z$ and $ c_3 =y \vee w \vee z$.} \label{Final}
\end{center}
\end{figure}

Without loss of generality suppose that $\Phi$ is a formula such that  every variable $x$, $x\in X$ appears in more than one clause. For every variable $x$, $x\in X$, we denote the number of clauses which contains $x$ by
$\gamma(x)$. For a given formula $\Phi$ we construct a 3-regular graph $G$, such that $\Phi$ has a NAE truth assignment if and only if $G$ has an
{\em  edge-labeling by sum} from $ \mathbb{N}_{2}$. For every variable $x$ if $\gamma(x)>2$, let $A_x$ be a cycle  of length $\gamma(x)$. Otherwise let $A_x$ be two vertices $a$ and $b$ with two parallel edges between $a$ and $b$.
Construct the graph $B_x$ by replacing every edge $ab$ of $A_x$ by $I(a,b)$ which is shown in Figure \ref{TBT}.
Let $c_{i_1}, \cdots, c_{i_{\gamma(x)}}$ be the clauses which contains $x$. $B_x$ has $\gamma(x)$
vertices of degree two, call these vertices $x_{i_1}, \cdots, x_{i_{\gamma(x)}}$, in an arbitrary order.
Now, for every variable $x$, $x\in X$, put a copy of $B_x$ and for every clause $c_i$, $c_i\in C$, put a vertex $c_i$.
For every clause $c_i$ and every variable $x$, if $x$ appears in $c_i$, then join $c_i$ to $x_{i}$.
Clearly, $G$ is a simple 3-regular graph (see Figure \ref{Final} for more details).

 First assume that $G$ has an {\em  edge-labeling $f$ by sum} from $ \mathbb{N}_{2}$ and let $\ell$ be the coloring which is induced by $f$. By  the structures of $A_x$ and $B_x$,  for a variable $x$ which appears in $c_{i_1}, \cdots, c_{i_{\gamma(x)}}$ the labels of edges $c_{i_1}x_{i_1}, \cdots, c_{i_{\gamma(x)}}x_{i_{\gamma(x)}}$ are the same.
Now, for every variable $x$, which appears in $c_{i_1}, \cdots, c_{i_{\gamma(x)}}$ put $\Gamma(x)=true$ if and only if the  label of edge $c_{i_1}x_{i_1}$  are 2.
Let $c_i$ be a clause with the variable $x$, $y$, $z$.
First, suppose that  $\{ f(x_i c_i), f(y_i c_i), f(z_i c_i)\}\neq\{1,2\}$, then $\ell(x)=\ell(c_i)$, but this is a contradiction.
So, we have $\{ f(x_i c_i), f(y_i c_i), f(z_i c_i)\}=\{1,2\}$.
Consequently, $\Gamma$ is a NAE satisfying assignment.
Next, suppose that $\Phi$ has a NAE satisfying assignment  $\Gamma : X  \rightarrow \{true, false\}$,
for every variable $x$, which  appears  in $c_{i_1}, \cdots, c_{i_{\gamma(x)}}$ label all the edges
$c_{i_1}x_{i_1}, \cdots, c_{i_{\gamma(x)}}x_{i_{\gamma(x)}}$ by 2 if and only if $\Gamma(x)=true$. It is easy to extent
this labeling to  an
{\em  edge-labeling by sum} from $ \mathbb{N}_{2}$. The proof is complete. $\clubsuit$

\section{Proof of Theorem \ref{pp1}}

{\bf (i)} Clearly, the problem is in $ \mathbf{NP} $. We reduce {\em  Cubic Planar 1-In-3 3-Sat} to our problem.
 Moore and Robson \cite{MR1863810} proved
that the following problem is $ \mathbf{NP}$-complete.

 {\em  Cubic Planar 1-In-3 3-Sat.}\\
\textsc{Instance}: A 3-Sat formula $\Phi=(X,C)$
 such that every variable
 appears in exactly three clauses, there
 is no negation in the formula, and the
bipartite graph obtained by linking a variable and a clause if and only
 if the
 variable appears in the clause, is planar.\\
\textsc{Question}: Is there a truth assignment for $X$ such that
 each clause in $C$ has exactly
one true literal?

\begin{figure}[ht]
\begin{center}
\includegraphics[scale=.52]{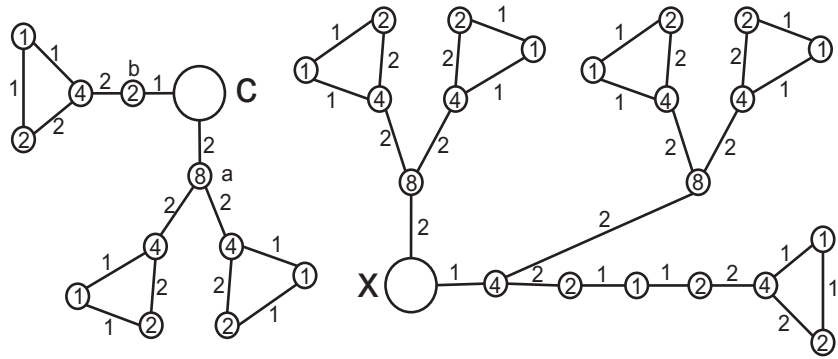}
\caption{The two gadgets $H_x$ and $I_c$. $I_c$ is on the left hand side of the figure.} \label{T1}
\end{center}
\end{figure}

Consider an instance $ \Phi $ of {\em  Cubic Planar 1-In-3 3-Sat}. We transform this into a  planar 3-colorable graph $G_{\Phi}$
such that $ G_{\Phi}$
has an
{\em  edge-labeling by product} from $ \mathbb{N}_{2}$
if and only if $\Phi$ has a 1-in-3  assignment.
 We use two gadgets $ H_x$ and $I_c$ which are shown in
Figure \ref{T1}. The graph $G_{\Phi}$ has
a copy of $H_x $ for each  variable $x \in X$ and  a copy of $I_c $ for each clause $c \in C$.
Also, for each clause $c =y \vee z \vee w$ add the edges $c y $, $c z  $ and $c w $.
First, suppose that $ G_{\Phi}$
has an
{\em  edge-labeling by product} from $ \mathbb{N}_{2}$.
In every copy of $ H_x$ and $I_c$ the label of every edge is determined. In Figure \ref{T1}, the label of each edge is written on the edge and the color of each vertex induced by edge labels is  written on the vertex. Every variable $x$
appears in
exactly three clauses, suppose that $x$ appears in $c_i$, $c_j$ and $c_k$.
By attention to the induced colors of neighbors of $x$ in $H_x$,
 the  labels of edges $c_i x$, $c_j x$ and $c_k x$ are equal.
Furthermore, by attention  to the $H_x$ and $I_c$, for every clause $c =x\vee y \vee z$, the  label of exactly one of the edges $c x$, $c y$ and $cz$ is 2 (note that if the labels of edges $c x$, $c y$ and $cz$ are 2, then the induced colors of $x$ and $c$ are same).
Now, for every variable $x$, which   appears in $c_i$, $c_j$ and $c_k$ put $\Gamma(x)=true$ if and only if the  labels of edges $c_i x$, $c_j x$ and $c_k x$ are 2. Clearly, $\Gamma$ is a 1-in-3 satisfying assignment.
Next, suppose that $\Phi$ has a 1-in-3 satisfying assignment  $\Gamma : X  \rightarrow \{true, false\}$,
for every variable $x$, which  appears in $c_i$, $c_j$ and $c_k$, label $c_i x$, $c_j x$ and $c_k x$ by 2 if and only if $\Gamma(x)=true$.
It is easy to extent this labeling
 to an {\em  edge-labeling by product} from $ \mathbb{N}_{2}$.\\ \\
{\bf (ii)} Let $G$ be a 3-regular graph. $G$ has an {\em  edge-labeling by product} from $ \mathbb{N}_{2}$ if and only if
$G$ has an {\em  edge-labeling by sum} from $ \mathbb{N}_{2}$, so by Theorem \ref{pp0}, we can prove the theorem.  $\clubsuit$

\section{Proof of Theorem \ref{pp2}}

{\bf (i)} Clearly, the problem is in $ \mathbf{NP}$. We reduced {\em  Cubic Planar 1-In-3 3-Sat} to our problem. In \cite{MR1863810} it was proved
that {\em  Cubic Planar 1-In-3 3-Sat} is $ \mathbf{NP}$-complete.
First, we construct an auxiliary graph $\mathcal{H}^c_i$. Put a copy of triangle $K_3=z_1^c z_2^c z_3^c$. For every vertex $z_j^c$, $1\leq j \leq 2 $, put $2i$ new isolated vertices $t_1^j,t_2^j, \ldots , t_{2i}^j$ and join $z_j^c$ to all of them. Also, add the edges $t_1^j t_2^j$, $t_3^j t_4^j, \ldots , t_{2i-1}^j t_{2i}^j$.  Next, put $2i-2$ new isolated vertices $t_1^3,t_2^3, \ldots , t_{2i-2}^3$ and join $z_3^c$ to all of them. Finally, add the edges $t_1^3 t_2^3$, $t_3^3 t_4^3, \ldots , t_{2i-3}^3 t_{2i-2}^3$.  Call the resulting graph $\mathcal{H}^{c}_i$.
Now, consider an instance $ \Psi $, we transform this into a planar 3-colorable  graph $G_{ \Psi}$
such that $ G_{\Psi}$
has a
{\em  vertex-labeling by product} from $ \mathbb{N}_{2}$
if and only if $\Psi$ has a 1-in-3  assignment.
 For each clause $c \in C$ put a vertex $c$ and a copy of $\mathcal{H}^{c}_3$, $\mathcal{H}^{c}_5$ and $\mathcal{H}^{c}_6$. Also, join the vertex $c$  to $z_3^c\in \mathcal{H}^{c}_3$, $z_3^c\in \mathcal{H}^{c}_5$ and $z_3^c n\in \mathcal{H}_{6}^{c}$.
 Next, for each  variable $x \in X$ put a vertex $x $. Finally,
for each clause $c =x \vee y \vee w$ add the edges $c x $, $c y  $ and $c w $.\\
First, suppose that $ G_{\Psi}$
has a
{\em  vertex-labeling $f$ by product} from $ \mathbb{N}_{2}$ and let $\ell$ be the induced coloring by $f$. In every copy of $\mathcal{H}_{3}^{c}$  the label of vertex $z_3^c$ is $2$. We have the similar property for  $\mathcal{H}_{5}^{c}$ and $\mathcal{H}_{6}^{c}$. By  the structure of $\mathcal{H}_{3}^{c}$, we have $f(c)=1$ and in the subgraph $\mathcal{H}_{3}^{c}$, $\ell(z_3^c)=8$; similarly for every subgraph $\mathcal{H}_{5}^{c}$,   $\ell(z_3^c)=32$ and for each subgraph $\mathcal{H}_{6}^{c}$, we have $\ell(z_3^c)=64$. So for every clause vertex $c$ we have $\ell(c)=16$.
Now, for every variable $x$,  put $\Gamma(x)=true$ if and only if $f(x)=2$.
Since for every clause $c$,   $\ell(c)=16$, $\Gamma$ is a 1-in-3  satisfying assignment.
On the other hand, suppose that $\Psi$ is 1-in-3 satisfiable with the satisfying assignment $\Gamma : X  \rightarrow \{true, false\}$,
for every variable $x$, label the vertex $x$ by 2 if and only if $\Gamma(x)=true$. The labels of other vertices are determined and it is clear that this labeling is a {\em  vertex-labeling by product} from $ \mathbb{N}_{2}$.\\ \\
{\bf (ii)} For every $k$, $k\geq 3$, we present a polynomial time reduction from {\em  3-colorability} to our problem.

{\em 3-Colorability}\\
\textsc{Instance}: A graph $G$.\\
\textsc{Question}: Is $\chi(G) \leq 3$?

First define the following sets: $   \mathcal{A}_k=\{ mn : m,n \in \mathbb{N} _k \}$, $\mathcal{B}_k = \{ \frac{m}{n} : m,n \in \mathbb{N}_k   \} $. Also,
define $\alpha(k)=\max_{\mathcal{D}_k \in \mathcal{C}_k} \mid \mathcal{D}_k \mid$, where $\mathcal{C}_k$ is the set of sets such that for every set $\mathcal{D}_k \in \mathcal{C}_k$, we have $\mathcal{D}_k \subseteq \mathcal{A}_k$ and $\{\frac{d}{d'}: d,d'\in \mathcal{D}_k \} \cap \mathcal{B}_k =\emptyset$. Since $k$ is constant, so we can compute $\alpha(k)$ in $O(1)$. Note that for $k$, $k\geq 3$, $\alpha(k)\geq 3$.
Now, for a given graph $G$ with $n$ vertices $v_1,v_2,\ldots, v_n$, join all vertices of $G$ to the all vertices of complete
graph $K_{\alpha(k)-3}$ with the vertices $v_{n+1},\ldots , v_{n+\alpha(k)-3}$. Call the resulting graph $G^*$.
Now consider the graph $G^{**}$ with the vertex set $\{ v_i^j : i\in \mathbb{N} _{n+\alpha(k)-3}, j\in \mathbb{N} _{k}\}$
such that $v_x^y$ is joined to $v_z^w$ if and only if $x=z$ or $v_x v_z \in E(G^*)$.
Finally, consider a copy of graph $G^{**}$, for every $i$, $1\leq i \leq n+\alpha(k)-3 $, put two new isolated vertices
$v_i'$ and $v_i''$ and join them to the set of vertices $\{v_i^1, \ldots, v_i^k \}$.
Call the resulting graph $\widetilde{G}$.

We show that $\widetilde{G}$ has a
{\em  vertex-labeling by product} from $ \mathbb{N}_{k}$ if and only if $G$ is 3-colorable.
Let $f$ be a {\em  vertex-labeling by product} for $\widetilde{G}$. Clearly, $f(v_1^1), \ldots , f(v_1^k)$ should be different integers. For every $i$, $i\in \mathbb{N} _{n+\alpha(k)-3}$, we have:
$\{ f(v_i^j): j \in \mathbb{N} _{k} \}= \mathbb{N} _{k}$.
Furthermore, for every $i_1$, $i_2$, $1\leq i_1 < i_2 \leq n+\alpha(k)-3$, we have:
$f(v_{i_1}')f(v_{i_1}''),f(v_{i_2}')f(v_{i_2}'') \in \mathcal{A}_k$. Also, for every $i_1$ and $i_2$, if
$v_{i_1}v_{i_2}\in E(G^*)$, then

$\dfrac{f(v_{i_1}')f(v_{i_1}'')}{f(v_{i_2}')f(v_{i_2}'')}\notin \mathcal{B}_k$.

Therefore,
$\mid \{ f(v_{i }')f(v_{i }''): 1\leq i \leq n+\alpha(k)-3 \} \mid \geq \alpha(k)-3 +\chi(G)$.
So, $\widetilde{G}$ has a
{\em  vertex-labeling by product} from $ \mathbb{N}_{k}$ if and only if $\chi(G)\leq 3$.
The proof is complete. $\clubsuit$

\section{Proof of Theorem \ref{random}}

For a given  $G(n,p)$, let $\{\mathcal{A}_1, \ldots , \mathcal{A}_5 \}$ be a partition of vertices to five parts such that
those parts are of equal or almost
equal sizes (that is, $\lfloor n/5 \rfloor $ or $ \lceil n/5 \rceil$).
For every vertex $v$ denote the number of neighbors of $v$ in $\mathcal{A}_i$ by $\lambda_v^i$.
Label the vertices of $ \mathcal{A}_1, \ldots , \mathcal{A}_5  $ by $2,3,5,7,11$, respectively.
This labeling is a {\em  vertex-labeling by product} if for every  two adjacent vertices $v$ and $u$, there is an index $i$, $1 \leq i \leq 5$ such that $\lambda_v^i \neq \lambda_u^i$.

\begin{eqnarray*}
\displaystyle Pr(\lambda_v^i = \lambda_u^i) & \leq &\Theta\Big( \sum_{t=0}^{  n/5  }\big( \Big( {  n/5   \atop t } \Big)p^t (1-p)^{  n/5 -t }\big)^2 \Big)\\
                                            &\leq & \Theta \Big(\max_{0\leq t \leq  n/5  }\Big( {  n/5 \atop t } \Big) p^t (1-p)^{  n/5  -t }\Big)\\
                                            &= & \Theta \Big( \Big( {  n/5   \atop   pn/5   } \Big) p^{p  n/5   } (1-p)^{(1-p)  n/5}\Big).
\end{eqnarray*}

By Stirling's approximation $\sqrt{2\pi n} (\frac{n}{e})^n \leq n! \leq \sqrt{e^2 n} (\frac{n}{e})^n$, we have:

\begin{eqnarray*}
         \displaystyle Pr(\lambda_v^i = \lambda_u^i)  & = &  \Theta(n^{-\frac{1}{2}}),\\
                        Pr(\forall i \ \lambda_v^i = \lambda_u^i)  & = &  \Theta(n^{-\frac{5}{2}}),\\
                      Pr(\exists vu \ \forall i \ \lambda_v^i = \lambda_u^i)  & = & \Theta(n^2) \Theta(n^{-\frac{5}{2}})=o(1).\\
\end{eqnarray*}

The proof is complete. $\clubsuit$

\section{Proof of Theorem \ref{pp3}}

{\bf (i)} Moret proved  in \cite{p} that  {\em Planar Not-All-Equal 3-Sat} is in $ \mathbf{P} $ by an interesting reduction to a known problem in $ \mathbf{P} $, namely Planar(Simple) MaxCut.

 {\em Planar Not-All-Equal 3-Sat.}\\
\textsc{Instance}: A 3-Sat formula $(X,C)$  such that the following graph obtained from 3-Sat is planar. The graph has one vertex for each variable, one vertex for each clause; all variable vertices are connected
in a simple cycle and each clause vertex is connected by an edge to variable
vertices corresponding to the literals present in the clause (note that positive
and negative literals are treated exactly alike).\\
\textsc{Question}: Is there a Not-All-Equal (NAE) truth assignment for $X$?

By a simple argument it was shown that the following problem is in  $ \mathbf{P} $ (for more information see \cite{ali}).

{\em Planar NAE 3-Sat Type 2.}\\
\textsc{Instance}: A 3-Sat formula $(X,C)$ such that  the following graph obtained from 3-Sat is planar. The graph has one vertex for each variable, one vertex for each clause and each clause vertex is connected by an edge to variable
vertices corresponding to the literals present in the clause.\\
\textsc{Question}: Is there a NAE truth assignment for $X$?\\
Now, consider the following variant.

 {\em Planar NAE Sat Type 2.}\\
\textsc{Instance}: Set $X$ of variables, collection $C$ of clauses over $X$ such that each
clause  $c \in C$ has $\mid c  \mid \geq 2$ and the following graph obtained from sat is planar. The graph has one vertex for each variable, one vertex for each clause and each clause vertex is connected by an edge to variable
vertices corresponding to the literals present in the clause.\\
\textsc{Question}: Is there a NAE truth assignment for $X$?

We can convert any instance  $\Phi$ of {\em Planar NAE Sat Type 2} to an instance $\Psi$ of
{\em Planar NAE 3-Sat Type 2} in polynomial time. For a given instance $\Phi$, for each clause with exactly two literals like $c=(x \vee y)$, put two clauses $x \vee y \vee t$ and $x \vee y \vee \neg t$ in $\Psi$, where $t$ is a new variable. Also, for each   clause with exactly four literals  $c=(x \vee y \vee w \vee z )$, put two clauses $x \vee y \vee t$ and $w \vee z \vee \neg t$ in $\Psi$, where $t$ is a new variable. For clauses with more than five variables we have a similar argument.

For a given graph $G$, we can use a depth-first search algorithm to identify the
connected components of $G$. $G$ has  an {\em edge-labeling  by gap} from $\{1,2\}$ if and only if each connected
component of $G$ has an {\em edge-labeling  by gap} from $\{1,2\}$.
So, without loss of generality we can assume that $G$ is connected.
Let $G[X,Y]$ be a connected planar bipartite graph such that $G$ does not have a vertex of degree one.
Let $f $ be an {\em edge-labeling  by gap} from $\{1,2\}$ for $G$.
The  color which is induced by $f$ for the set of vertices $X$ is one and the induced color by $f$ for the set of vertices $Y$ is zero, or vice versa ({{\bf Property 1}}).

For every vertex $v\in X$, consider a variable $v$ in $\Phi$ and for every vertex $u\in Y$ put a clause $(\bigvee_{v\sim u} v)$ in $\Phi$. Now determine whether $\Phi$  has a NAE truth assignment.
If $\Phi$  has a NAE truth assignment $\Gamma$, for every vertex $v$, $v\in X$ label all edge incident with  $v$ by $1$ if and only if
$\Gamma(v)=false$. Label other edges of $G$ by $2$, call this labeling  $f$. It is easy to see that $f$ is an {\em   edge-labeling  by gap}.
If  $\Phi$  does not have a NAE truth assignment. Then, for every vertex $v\in Y$, consider a variable $v$ in $\Psi$ and for every vertex $u\in X$ put a clause $(\bigvee_{v\sim u} v)$ in $\Psi$. Now determine whether $\Psi$  has a NAE truth assignment. If $\Psi$  has a NAE truth assignment $\Gamma$ by a similar method we can find an
{\em   edge-labeling $f$ by gap} from $\{1,2\}$ for $G$. If $\Psi$  does not have a NAE truth assignment, by Property 1, $G$ does not have any {\em   edge-labeling by gap} from $\{1,2\}$.\\ \\
{\bf (ii)} Let $\Phi$ be a $3$-Sat formula with clauses $C=\lbrace
c_1, \ldots ,c_k\rbrace $ and variables
$X=\lbrace x_1, \ldots ,x_n\rbrace $. Let $G(\Phi)$ be a graph with the vertices $C \cup X \cup (\neg X)$, where $\neg X = \lbrace \neg x_1, \ldots , \neg x_n\rbrace$, such that for each clause $c_j=y \vee z \vee w $, $c_j$ is adjacent to $y,z$ and $w$, also every $x_i \in X$ is adjacent to $\neg x_i$. $\Phi$ is called planar $3$-Sat type $2$  formula if $G(\Phi)$ is a planar graph. It was shown that the problem of satisfiability of planar $3$-Sat type $2$  is $ \mathbf{NP}$-complete \cite{zhu1}.

{\em   Planar $3$-Sat type $2$.}\\
\textsc{Instance}: A $3$-Sat type $2$  formula  $ \Phi $.\\
\textsc{Question}: Is there a truth assignment for $ \Phi $ that satisfies all the clauses?

In {\em planar $3$-Sat type $2$}, if we only consider the set of formulas  $\mathcal{F}$ such that
for every $\Phi$, $\Phi \in \mathcal{F}$, $G(\Phi)$   is connected, the problem remains
$ \mathbf{NP} $-complete. We reduce this version to our problem.
Consider an instance $ \Phi $, we transform this into a  graph $G_{\Phi}$
such that $ G_{\Phi}$
has an
{\em  edge-labeling by gap} from $ \mathbb{N}_{2}$
if and only if $\Phi$ has a satisfying  assignment.
For each  variable $x \in X$  put a a copy of cycle $C_4=x t_x \neg x   t_{x}'$. For each clause $c \in C$ put a copy of  gadget $P_4=c c' c'' c'''$. Now, put a copy $C_6$.
Also, for each clause $c =y \vee z \vee w$ add the edges $c y $, $c z  $ and $c w $.
Finally, let $x$ be an arbitrary variable, join $x$ to one of the vertices of $C_6$.
Call the resulting graph $ G_{\Phi}$.
$ G_{\Phi}$ is connected, bipartite and planar.
First, suppose that $ G_{\Phi}$
has an
{\em  edge-labeling $f$ by gap} from $ \mathbb{N}_{2}$ and $l$ is the  proper coloring which is induced by $f$.
Since for every variable $x$ the degrees of vertices $x$ and $ \neg x $ are greater than one, also for every clause $c$ the degree of vertex $c $ is 4 and $ G_{\Phi}$ is connected, hence in the induced coloring $l$ by $f$, for the set of variables and the set of clauses, we have $l(x_1)=l(\neg x_1)=\ldots=l(\neg x_n)=l(x_n)\neq l(c_1)=\ldots=l(c_m)$ and $l(x_1)\neq 2 \neq l(c_1)$.

First, suppose that $l(x)=1$, since $x$ is adjacent to one of the vertices of $C_6$, in this situation $ G_{\Phi}$
does not have any
{\em  edge-labeling $f$ by gap} from $ \mathbb{N}_{2}$. So for all $x\in X$ and $c\in C$, we have $l(x)=0$, $l(\neg x)=0$ and $l(c)=1$. Hence, the labels of all edges incident with $x$ are same. Also, for every variable $x$, because of $t_x$, the labels of all edges incident with $x$ are different from
the labels of all edges incident with $\neg x$.
Now,  for every variable $x$, which  appears in $c_i, c_j, \ldots , c_k$ put $\Gamma(x)=true$ if and only if the  label of edge $c_i x$ is $2$.
For every clause $c =x \vee y \vee w$, $l(c)=1$. If the  labels of  all edges $  cx, cy, cw $ are  1, then
$f(cc')=2$. Thus $f(c'c'')=2 $ and $f(c''c''')=1$, so $l(c'')=l(c''')$, but this is a contradiction. Therefore,
 $2 \in \{ f(cx),f(cy),f(cw)\}$.
 Therefore, $\Gamma$ is an satisfying assignment.
Now, let $\Gamma$ be a satisfying assignment for $\Phi$. For every variable $x$, label all the edges incident with $x$ by $2$ if and only if $\Gamma(x)=true$. It is easy to extend this labeling to an {\em  edge-labeling $f$ by gap} from $\{1,2 \}$.
This completes the proof.\\ \\
{\bf (iii)} For every $k$, we present a polynomial time reduction from {\em  k-colorability}, $k>2$, to our problem.

{\em k-Colorability}\\
\textsc{Instance}: A graph $G$.\\
\textsc{Question}: Is $\chi(G) \leq k$?

For a given connected graph $G$ with more than two vertices, we construct a graph $G^*$ such that $\chi(G) \leq k$ if and only if $G^*$ has an
{\em  edge-labeling by gap} from $ \mathbb{N}_{k}$.
For every vertex $v$, $v \in V(G )$, put a copy of gadget $P_3=v'v''v'''$
and join $v'$ to $u'$ if and only if $vu\in E(G )$. Call the resulting graph  $G^{*}$.
First, suppose that $G^{*}$ has an
{\em  edge-labeling $f$ by gap} from $ \{ 1,2,\cdots , k\}$ and $\ell$ is the  coloring which is induced by $f$. For every
vertex $v'$, we have $\ell(v')\in \{0,1,\ldots, k-1\}$, so $\ell$ is  a proper vertex coloring for $G$. Now, let $c$ be a proper vertex coloring for $G$. For every vertex $v'$, $v'\in V(G^*)$, label
all edges incident with $v'$ except $v'v''$ by $1$ and label  $v'v''$ by $c(v)$.
Finally for every edge $v''v'''$, label $v''v'''$ by $c(v)$ if $c(v)\neq 1$, otherwise label $v''v'''$ by $k$. This labeling
is an {\em  edge-labeling by gap} from $ \mathbb{N}_{k}$.
$\clubsuit$

\section{Proof of Theorem \ref{pp4}}

\begin{algorithm}
\caption{(Vertex Labeling By Gap From $\mathbb{N}_{2}$)}
\begin{algorithmic}[1]
\STATE {{\bf Input:}} A  planar bipartite graph $G[X,Y]$
\STATE {{\bf Question:}} Does $G$ have a {\em  vertex-labeling by gap} from $\mathbb{N}_{2}$?
\STATE  Let $G_1, \cdots , G_d$ be the connected components of $G$.
   \STATE  $j\leftarrow 0$
\FOR{ $i=1$ to $i=d$}
\STATE $G_1'[X',Y'] \leftarrow G_1[X,Y]$
\STATE Remove all vertices of degree one from $G_1'$
\STATE $\Phi \leftarrow \emptyset$
\FOR{  every vertex $u\in V(Y')$ }
         \STATE $\Phi \leftarrow \Phi \cup (\bigvee_{vu\in E(G)} v)$
\ENDFOR
\STATE $\Psi \leftarrow \emptyset$
\FOR{  every vertex $v\in V(X')$ }
         \STATE $\Psi \leftarrow \Psi \cup (\bigvee_{vu\in E(G)} u)$
\ENDFOR
\IF{ $ \Phi$ has a Not-All-Equal truth assignment}
   \STATE  $j\leftarrow j+1$
\ELSIF{ $ \Psi$ has a Not-All-Equal truth assignment}
    \STATE  $j\leftarrow j+1$
\ENDIF
\ENDFOR
\IF{ j=d}
   \RETURN  $G$ has a vertex-labeling by gap from $ \mathbb{N}_{2}$.
\ENDIF
   \RETURN  $G$ does not have any vertex-labeling by gap from $ \mathbb{N}_{2}$.
\end{algorithmic}
\end{algorithm}

{\bf (i)}
For a given graph $G$, we can use a depth-first search algorithm to identify the
connected components of $G$. $G$ has  a {\em vertex-labeling  by gap} from $\{1,2\}$ if and only if each connected
component of $G$ has a  {\em vertex-labeling  by gap} from $\{1,2\}$.
So, without loss of generality we can assume that $G$ is connected.
Let $G[X,Y]$ be a connected planar bipartite graph with more than two vertices. Consider a copy of $G$, remove all vertices of degree one from $G$ and
call the resulting graph $G'[X',Y']$, where $X'\subseteq  X$. If $E(G')=\emptyset$, then $G$ is a star, so $G$ has a {\em   vertex-labeling  by gap} from $\{1,2\}$.
So suppose that $E(G')\neq\emptyset$ and let $f $ be a {\em   vertex-labeling  by gap} from $\{1,2\}$ for $G$.
Since $G'[X',Y']$ is  connected  and bipartite, so the  color which is induced by $f$ for the set of vertices $X'$ is one and the  color which is induced by $f$ for the set of vertices $Y'$ is zero, or vice versa ({{\bf Property A}}).

For every vertex $v\in X$, consider a variable $v$ in $\Phi$ and for every vertex $u\in Y'$ put a clause $(\bigvee_{vu\in E(G)} v)$ in $\Phi$. Now determine whether $\Phi$  has a NAE truth assignment.
If $\Phi$  has a NAE truth assignment $\Gamma$, for every vertex $v$, $v\in X$ label $v$ by $1$ if and only if
$\Gamma(v)=false$. Label other vertices of $G$ by $2$.
It is easy to see that this labeling is a {\em   vertex-labeling by gap} from $\{1,2\}$ for $G$.
If  $\Phi$  does not have a NAE truth assignment, then, for every vertex $v\in Y$, consider a variable $v$ in $\Psi$ and for every vertex $u\in X'$  put a clause $(\bigvee_{vu\in E(G)} v)$ in $\Psi$. Now determine whether $\Psi$  has a NAE truth assignment. If $\Psi$  has a NAE truth assignment $\Gamma$ by a similar method we can find
{\em   vertex-labeling by gap} from $\{1,2\}$ for $G$. Otherwise, by Property $A$, $G$ does not have any {\em   vertex-labeling by gap} from $\mathbb{N}_{2}$.  See Algorithm 1.\\

{\bf (ii)} We reduce {\em   NAE 3-Sat} to our problem in polynomial time.
For a given $\Phi$, we transform $\Phi$ into a  graph $G_{\Phi}$
such that $ G_{\Phi}$
has a
{\em vertex-labeling by gap} from $ \mathbb{N}_{2}$
if and only if $\Phi$ has a NAE satisfying  assignment.
Construction of $ G_{\Phi}$ is similar to the
construction of $ G_{\Phi}$ in the proof of (iii)
of Theorem \ref{pp3}, except the gadget $P_4=cc'c''c'''$. For each clause $c \in C$ instead of $P_4=cc'c''c'''$, put an isolated vertex  $c $.
First, suppose that $ G_{\Phi}$
has a
{\em vertex-labeling $f$ by gap} from $ \mathbb{N}_{2}$ and $l$ is the induced proper coloring by $f$. By an argument similar to argument of proof of Theorem \ref{pp3}, for every clause $c =x \vee y \vee w$, $l(c)=1$.
So $\{f(x),f(y),f(w)\}=\{1,2\}$, therefore $\Gamma$ is a NAE satisfying assignment.
Now, let $\Gamma$ be an satisfying assignment for $\Phi$. For every variable $x$, label the vertex $x$ by $2$ if and only if $\Gamma(x)=true$. This completes the proof.\\ \\
{\bf (iii)} The proof is similar to the proof of part (iii) of Theorem \ref{pp3}.\\ \\
{\bf (iv)} It was shown that   3-colorability of planar 4-regular
graphs  is NP-complete \cite{MR573644}. By a proof similar to the proof of part (iii) of Theorem \ref{pp3}, we can prove the theorem.
$\clubsuit$

\section{Proof of Theorem \ref{pp5}}
{\bf (i)} We reduce our problem to 2-Sat problem in polynomial time.

{\em  2-Sat.}\\
\textsc{Instance}: A 2-Sat formula $\Phi$.\\
\textsc{Question}: Is there a truth assignment for $\Phi$ that satisfies all the clauses?

For a given graph $G$ of order $n$ we construct a 2-Sat formula $ \Phi$ with $n$ variables $v_1, \ldots , v_n$
such that $G$ has a vertex-labeling by degree from $\{1, 2\}$ if and only if there is a truth assignment for $\Phi$. For every edge $e = v_iv_j$ , if $d(v_i) = d(v_j)$, add the clauses $v_i\vee v_j$ and $\neg v_i \vee \neg v_j$
and if $d(v_i) = 2d(v_j)$, add the clause $v_i\vee  \neg v_j$ , otherwise if $2d(v_i) = d(v_j)$, add the clause
$\neg v_i \vee v_j $. First, suppose that $\Gamma$ is a satisfying assignment for $\Phi$. For every vertex $v_i$, label $v_i$
by 2 if and only if $\Gamma(v_i) = true$. It is easy to see that this labeling is a vertex-labeling by
degree from $\{1, 2\}$. Next, let $f$ be a {\em  vertex-labeling by degree} from $\{1, 2\}$, for every variable
$v_i$, put $\Gamma(v_i) = true$ if and only if $f(v_i) = 2$. As we know, 2-Sat problem is in P \cite{MR1567289}.
This completes the proof.\\ \\
{\bf (ii)} It was shown that   3-colorability of planar 4-regular
graphs  is NP-complete \cite{MR573644}. For a given 4-regular
graph $G$ and a number $k$, $k\geq 3$, we construct a  graph $\mathcal{H}^*$ with chromatic number $k+1$ in linear time such that
$\mathcal{H}^*$  has a {\em  vertex-labeling by degree} from $\mathbb{N}_{k}$ if and only if $G$ is 3-colorable.
Let $\mathcal{P}_k=\{4,5,\ldots,k\}$.

{{\bf  Construction of $\mathcal{H}^*$.}} First, consider a copy of graph $G$ and call its vertices old vertices . For every old vertex $v$  and for every $\alpha \in \mathcal{P}_k $, consider two copies of complete graphs $K_{k-1}$. Call this two copies by $K_{k-1}^{(v,\alpha)}$ and $K_{k-1}^{(v,\alpha)'}$. Join $v$ to all vertices of $K_{k-1}^{(v,\alpha)}$ and $K_{k-1}^{(v,\alpha)'}$, Next join all
vertices of $K_{k-1}^{(v,\alpha)}$ to a new vertex $v_{\alpha}$ and also join all
vertices of $K_{k-1}^{(v,\alpha)'}$ to a new vertex $v_{\alpha}'$. Now, join $v_{\alpha}$ to $v_{\alpha}'$.
Call the resulting graph $\mathcal{H}$.
For a given graph $\mathcal{H}$, for every old vertex $v$ and for every $\alpha \in \mathcal{P}_k $, join every vertex
of $u$, $u\in V(K_{k-1}^{(v,\alpha)})\cup V(K_{k-1}^{(v,\alpha)'})\cup \{ v_{\alpha},v_{\alpha}' \}$ to $\alpha \times d_{\mathcal{H}}(v)- d_{\mathcal{H}}(u)$ new isolated vertices $l_u^{1}, l_u^{2}, \ldots, l_u^{\alpha \times d_{\mathcal{H}}(v)- d_{\mathcal{H}}(u)} $. Call the resulting $\mathcal{H}^*$.

The chromatic number of $\mathcal{H}^*$ is $k+1$. Because every two vertices in $V(K_{k-1}^{(v,\alpha)})\cup \{ v_{\alpha}\}$ have the same degree,  every {\em vertex-labeling by degree} needs at
least $k$ distinct labels. Let $f$ be a  {\em vertex-labeling by degree} from $\{1,\ldots,k\}$ for $\mathcal{H}^*$.
We claim that for every old vertex $v$, $f(v)\in \{1,2,3\}$.
To the contrary suppose that $f(v)\geq 4$, so there is a $\alpha \in \mathcal{P}_k$ such that $\alpha= f(v)$. Therefore,
there is no vertex in $K_{k-1}^{(v,\alpha)}$ with the label $1$, hence $l(v_{\alpha})=1$. Similarly, $l(v_{\alpha}')=1$. But
$v_{\alpha}$ and $v_{\alpha}'$ are adjacent and have the same degree. This is a contradiction.
So, for every old vertex $v$, we have $f(v)\in \{1,2,3\}$. Since the degree of old vertices are equal, therefore $f$ is a proper vertex coloring for old vertices, so $G$ is 3-colorable. On the other hand,
let $c$ be proper vertex coloring for $G$ with 3 colors. It is easy to find a   {\em vertex-labeling by degree} from $\{1,\cdots,k\}$ for $\mathcal{H}^*$.
$\clubsuit$

\section{Proof of Theorem \ref{pp6}}

\begin{figure}[ht]
\begin{center}
\includegraphics[scale=.42]{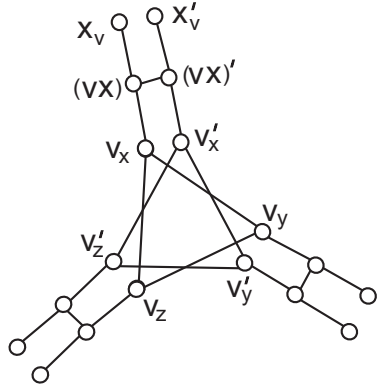}
\caption{Transformation which is used in constructing $G'$.} \label{graphA}
\end{center}
\end{figure}

{\bf (i)} Clearly, the problem is in $ \mathbf{NP} $. It was shown that it is $ \mathbf{NP} $-hard to determine the edge chromatic number of a
cubic graph \cite{MR635430}. Let $G$ be a $3$-regular graph. We construct a $3$-regular graph $G'$ from $G$ such that $G'$ has a
{\em  vertex-labeling by maximum} from $ \mathbb{N}_{3}$ if and only if $G$ belongs to $Class$ $1$. In order to construct $G'$, for every vertex $v\in V(G)$ with the neighbors $x$, $y$ and $z$ consider two disjoint triangles $v_x v_y v_z$ and $v_{x}' v_{y}' v_{z}'$ in $G'$. Also, for every edge $e\in E(G)$, consider two adjacent vertices $e$ and $e'$ in $G'$. Finally, for every edge $e=uv \in E(G)$, join $e$ to $v_u$ and $u_v$; also join $e'$ to $v_{u}'$ and $u_{v}'$. Name the constructed graph $G'$ (see Figure \ref{graphA}).
Since $G'$ has triangles, so
every
{\em  vertex-labeling by maximum} needs at least $3$ distinct labels.
First suppose that $G'$ has a
{\em  vertex-labeling  $f $ by maximum} from $ \mathbb{N}_{3}$ and let $\ell$ be the induced vertex coloring by $f$. For every vertex $v\in V(G)$ with the neighbors $x$, $y$ and $z$ in $G$, we have $\{\ell( v_x), \ell( v_y), \ell(v_z)\}=\{1,2,3\}=\{ \ell(v_{x}') ,\ell(v_{y}'), \ell(v_{z}')\}$. Suppose that there are $u$ and $v$ such that
$\ell( v_u) = \ell(v_{u}')=3$,  since $f$ can not assign $3$ to a vertex in a triangle, so  $f(vu)=f((vu)')=3$. Hence $\ell(vu)= \ell( (vu)')=3$ and this is a contradiction. So we have the following fact:

There are no $u$ and $v$ such that $\ell(v_u) = \ell( v_{u}') =3$ ({{\bf Fact 1}}).

Now, consider the following proper $3$-edge coloring for $G$.

$g: E(G) \longrightarrow \lbrace 1,2 , 3 \rbrace$,\\
$g(uv)=
\begin{cases}
   1       & $if$\,\,f(uv)=3,\\
   2       & $if$\,\,f((uv)')=3,\\
   3       & $otherwise$.\
\end{cases}$

By Fact 1, $g$ is well-defined and $G$ belongs to $Class$ $1$. On the other hand, assume that $g: E(G) \longrightarrow \lbrace 1,2 , 3 \rbrace$ is a proper 3-edge coloring. Define $f: V(G') \longrightarrow \lbrace 1,2 , 3 \rbrace$ such that for every edge $uv\in E(G)$, $f(v_u)=f(v_{u}')=1$, $f(uv)=g(uv)$ and $f((uv)') \equiv g(uv)+1(\mod 3)$. It is easy to see that $f$ is a {\em  vertex-labeling by maximum}.\\ \\
{\bf (ii)}
First, consider the following simple procedure:

$ \diamond$  For a given graph $G$, put a new vertex $v$ and join it to the all vertices of $G$, next put a new vertex $u$ and join it to $v$. Name the constructed graph $G'$.

 We can construct  $G'$  in polynomial time and $G$ has a
{\em  vertex-labeling by maximum} from $ \mathbb{N}_{k}$
if and only if $G'$ has a
{\em  vertex-labeling by maximum} from $ \mathbb{N}_{k+1}$.
In Part (i), we proved that, it is $ \mathbf{NP} $-complete to decide  whether a given $3$-colorable  graph $G$ has a
{\em  vertex-labeling by maximum} from $ \mathbb{N}_{3}$. So by repeating the above procedure, for every
$k$, $k\geq 3$, we can prove that, it is $ \mathbf{NP} $-complete to decide  whether a given $k$-colorable graph $H$ has a
{\em  vertex-labeling by maximum} from $ \mathbb{N}_{k}$.
This completes the proof.

\begin{algorithm}
\caption{(TSV Finder)}
\begin{algorithmic}[1]
\STATE {{\bf Input:}} A   graph $G$
\STATE {{\bf Question:}} Does $G$ have a {\em  vertex-labeling by maximum}?
\STATE $S \leftarrow \emptyset$
  \FOR{  every vertex $u$ appears in a triangle }
  \STATE $S\leftarrow S \cup \{ u \}$
 \ENDFOR
\WHILE{ there are two adjacent vertices $u$ and $v$ such that $v\in S$, $u \in N(S)\setminus S$ and there is no  vertex $z\in S$ such that $z$ is adjacent to $v$ and $u$  {{\bf or}}  $v$ does not appear in any triangle in $G[S]$}
\STATE $S \leftarrow S \setminus \{ v\}$
\ENDWHILE
\IF{ $S \neq \emptyset $}
\RETURN $G$ has the TSV $S$.
\ELSE
\RETURN  $G$ does not have a TSV.
\ENDIF
\end{algorithmic}
\end{algorithm}

{\bf (iii)}
Consider Algorithm $2$. When Algorithm 2 terminates,
if it returns "$G$ has the TSV $S$", then $S$ is a TSV, so $G$ does not have any {\em  vertex-labeling by maximum}.
Suppose that Algorithm $2$ returns "$G$ does not have a TSV", but $G$ has a TSV $S'$. In the lines $2$ and $3$ of the algorithm, the set of vertices $S'$ are added to $S$. Now, consider the line $5$ of the algorithm and let $v\in S'$ be the first vertex from the set $S'$ that is eliminated from $S$. When Algorithm $2$ chooses the vertex $v$, $v$ is in a triangle in $G[S']$, so it is in a triangle in $G[S]$. Therefore, there is  a vertex $u$  such that $uv\in E(G)$, $v\in S'$, $u \in N(S)\setminus S$ and there is no  vertex $z\in S$ such that $z$ is adjacent to $v$ and $u$. So $S'$ is not a TSV. This is a contradiction. So when Algorithm $2$ returns "$G$ has no TSV", $G$ does not have any TSV. $\clubsuit$

\section{Acknowledgment}
We would like to thank Wiktor \.{Z}elazny  for his valuable answers to our questions about the definition of fictional coloring. Also, we would like to express our deep gratitude to the referees for their constructive and fruitful comments.

\bibliographystyle{plain}
\bibliography{luckyref}

\begin{thebibliography}{10}

\bibitem{MR2145514}
L.~Addario-Berry, R.~E.~L. Aldred, K.~Dalal, and B.~A. Reed.
\newblock Vertex colouring edge partitions.
\newblock {\em J. Combin. Theory Ser. B}, 94(2):237--244, 2005.

\bibitem{MR2404230}
L.~Addario-Berry, K.~Dalal, and B.~A. Reed.
\newblock Degree constrained subgraphs.
\newblock {\em Discrete Appl. Math.}, 156(7):1168--1174, 2008.

\bibitem{ahadi}
A.~Ahadi, A.~Dehghan, M.~Kazemi, and E.~Mollaahmadi.
\newblock Computation of lucky number of planar graphs is {NP}-hard.
\newblock {\em Inform. Process. Lett.}, 112(4):109--–112, 2012.

\bibitem{akbari2}
S.~Akbari, M.~Ghanbari, R.~Manaviyat, and S.~Zare.
\newblock On the lucky choice number of graphs.
\newblock {\em Graphs Combin.}, (to appear).

\bibitem{PNT}
Eric Bach and Jeffrey Shallit.
\newblock {\em Algorithmic Number Theory}.
\newblock MIT Press, 1996.

\bibitem{MR2299707}
P.~N. Balister, E.~Gy{\H{o}}ri, J.~Lehel, and R.~H. Schelp.
\newblock Adjacent vertex distinguishing edge-colorings.
\newblock {\em SIAM J. Discrete Math.}, 21(1):237--250, 2007.

\bibitem{z2}
T.~Bartnicki, J.~Grytczuk, and S.~Niwczyk.
\newblock Weight choosability of graphs.
\newblock {\em J. Graph Theory}, 60(3):242--256, 2009.

\bibitem{MR1469354}
A.~C. Burris and R.~H. Schelp.
\newblock Vertex-distinguishing proper edge-colorings.
\newblock {\em J. Graph Theory}, 26(2):73--82, 1997.

\bibitem{MR2729020}
Gary Chartrand, Futaba Okamoto, and Ping Zhang.
\newblock The sigma chromatic number of a graph.
\newblock {\em Graphs Combin.}, 26(6):755--773, 2010.

\bibitem{MR2552893}
Sebastian Czerwi{\'n}ski, Jaros{\l}aw Grytczuk, and Wiktor {\.Z}elazny.
\newblock Lucky labelings of graphs.
\newblock {\em Inform. Process. Lett.}, 109(18):1078--1081, 2009.

\bibitem{MR573644}
David~P. Dailey.
\newblock Uniqueness of colorability and colorability of planar {$4$}-regular
  graphs are {NP}-complete.
\newblock {\em Discrete Math.}, 30(3):289--293, 1980.

\bibitem{ali}
A.~Dehghan, M.-R. Sadeghi, and A.~Ahadi.
\newblock The complexity of the sigma chromatic number of cubic graphs.
\newblock {\em Submitted to Discrete Appl. Math.}

\bibitem{zhu1}
Ding-Zhu Du, Ker-K Ko, and J.~Wang.
\newblock {\em Introduction to Computational Complexity}.
\newblock Higher Education Press, 2002.

\bibitem{David}
Andrzej Dudek and David Wajc.
\newblock On the complexity of vertex-coloring edge-weightings.
\newblock {\em Discrete Math. Theor. Comput. Sci.}, 13(3):45--50, 2011.

\bibitem{MR1567289}
M.~R. Garey and D.~S. Johnson.
\newblock {\em {C}omputers and intractability: {A} guide to the theory of
  {$NP$}-completeness}.
\newblock W. H. Freeman, San Francisco, 1979.

\bibitem{MR2171364}
Hamed Hatami.
\newblock {$\Delta+300$} is a bound on the adjacent vertex distinguishing edge
  chromatic number.
\newblock {\em J. Combin. Theory Ser. B}, 95(2):246--256, 2005.

\bibitem{MR635430}
Ian Holyer.
\newblock The {NP}-completeness of edge-coloring.
\newblock {\em SIAM J. Comput.}, 10(4):718--720, 1981.

\bibitem{MR2595676}
Maciej Kalkowski, Micha{\l} Karo{\'n}ski, and Florian Pfender.
\newblock Vertex-coloring edge-weightings: towards the 1-2-3-conjecture.
\newblock {\em J. Combin. Theory Ser. B}, 100(3):347--349, 2010.

\bibitem{MR2047539}
Micha{\l} Karo{\'n}ski, Tomasz {\L}uczak, and Andrew Thomason.
\newblock Edge weights and vertex colours.
\newblock {\em J. Combin. Theory Ser. B}, 91(1):151--157, 2004.

\bibitem{z3}
M.~Khatirinejad, R.~Naserasr, M.~Newman, B.~Seamone, and B~Stevens.
\newblock Digraphs are 2-weight choosable.
\newblock {\em Electron. J. Combin.}, 18(1):Paper 21,4, 2011.

\bibitem{D2}
M.~Khatirinejad, R.~Naserasr, M.~Newman, B.~Seamone, and B.~Stevens.
\newblock Vertex-colouring edge-weightings with two edge weights.
\newblock {\em Discrete Math. Theor. Comput. Sci.}, 14(1):1--20, 2012.

\bibitem{MR1863810}
C.~Moore and J.~M. Robson.
\newblock Hard tiling problems with simple tiles.
\newblock {\em Discrete Comput. Geom.}, 26(4):573--590, 2001.

\bibitem{p}
B.~M. Moret.
\newblock Planar {NAE3SAT} is in {P}.
\newblock {\em SIGACT News 19, 2}, pages 51--54, 1988.

\bibitem{MR2428690}
Joanna Skowronek-Kazi{\'o}w.
\newblock 1,2 conjecture---the multiplicative version.
\newblock {\em Inform. Process. Lett.}, 107(3-4):93--95, 2008.

\bibitem{product}
Joanna Skowronek-Kazi{\'o}w.
\newblock Multiplicative vertex-colouring weightings of graphs.
\newblock {\em Inform. Process. Lett.}, 112(5):191--–194, 2012.

\bibitem{gap}
M.A. Tahraoui, E.~Duchene, and H.~Kheddouci.
\newblock Gap vertex-distinguishing edge colorings of graphs.
\newblock {\em Discrete Math.}, 312(20):3011--–3025, 2012.

\bibitem{MR1135027}
Carsten Thomassen.
\newblock The even cycle problem for directed graphs.
\newblock {\em J. Amer. Math. Soc.}, 5(2):217--229, 1992.

\bibitem{MR0180505}
V.~G. Vizing.
\newblock On an estimate of the chromatic class of a {$p$}-graph.
\newblock {\em Diskret. Analiz No.}, 3:25--30, 1964.

\bibitem{MR1367739}
Douglas~B. West.
\newblock {\em Introduction to graph theory}.
\newblock Prentice Hall Inc., Upper Saddle River, NJ, 1996.

\bibitem{personal}
Wiktor \.{Z}elazny.
\newblock {\em Personal Communication}.

\end{thebibliography}

\end{document}